\documentclass[twoside,twocolumn]{article}

\usepackage{blindtext} 

\usepackage[sc]{mathpazo} 
\usepackage[T1]{fontenc} 
\usepackage{microtype} 

\usepackage[english]{babel} 

\usepackage[hmarginratio=1:1,top=32mm,columnsep=20pt]{geometry} 
\usepackage[hang, small,labelfont=bf,up,textfont=it,up]{caption} 
\usepackage{booktabs} 

\usepackage{lettrine} 

\usepackage{enumitem} 
\setlist[itemize]{noitemsep} 

\usepackage{abstract} 

\usepackage{titlesec} 
\renewcommand\thesection{\Roman{section}} 
\renewcommand\thesubsection{\roman{subsection}} 
\titleformat{\section}[block]{\large\scshape\centering}{\thesection.}{1em}{} 
\titleformat{\subsection}[block]{\large}{\thesubsection.}{1em}{} 

\usepackage{fancyhdr} 
\usepackage{color}
\usepackage{url}
\usepackage{cite}
\usepackage[usenames,dvipsnames]{pstricks}
\usepackage{graphicx}
\usepackage[cmex10]{amsmath}

\usepackage{array}
\usepackage{multirow}
\usepackage{algorithmic}
\usepackage{caption}
\usepackage{subcaption}

\newcommand\Tau{\mathcal{T}}
\usepackage{titling} 

\usepackage{hyperref} 

\pretitle{\begin{center}\Huge\bfseries} 
\posttitle{\end{center}} 
\title{Close Miking Empirical Practice Verification: A Source Separation Approach} 
\author{%
\textsc{Konstantinos Drossos} \\ 
\normalsize Audio Research Group,\\ \normalsize Dept. of Signal Processing,\\ \normalsize Tampere University of Technology,\\ \normalsize Tampere, Finland. \\ 
\and 
\textsc{Stylianos Ioannis Mimilakis}\thanks{Correspondance should be addressed to {mis@idmt.fraunhofer.de}} \\ 
\normalsize Fraunhofer IDMT,\\ \normalsize Ilmenau, Germany. \\ 
\and
\textsc{Andreas Floros} \\ 
\normalsize Lab of Audiovisual Signal Processing,\\ \normalsize Dept. of Audiovisual Arts,\\ \normalsize Ionian University, Corfu, Greece. \\ 
\and 
\textsc{Tuomas Virtanen} \\ 
\normalsize \normalsize Audio Research Group,\\ \normalsize Dept. of Signal Processing,\\ \normalsize Tampere University of Technology,\\ \normalsize Tampere, Finland. \\ 
\and
\textsc{Gerald Schuller} \\ 
\normalsize Technical University of Ilmenau,\\ \normalsize Ilmenau, Germany. \\ 
}
\date{} 
\renewcommand{\maketitlehookd}{%
\begin{abstract}
\noindent Close miking represents a widely employed practice of placing a microphone very near to the sound source in order to capture more direct sound and minimize any pickup of ambient sound, including other, concurrently active sources. It is used by the audio engineering community for decades for audio recording, based on a number of empirical rules that were evolved during the recording practice itself. But can this empirical knowledge and close miking practice be systematically verified? In this work we aim to address this question based on an analytic methodology that employs techniques and metrics originating from the sound source separation evaluation field. In particular, we apply a quantitative analysis of the source separation capabilities of the close miking technique. The analysis is applied on a recording dataset obtained at multiple positions of a typical musical hall, multiple distances between the microphone and the sound source multiple microphone types and multiple level differences between the sound source and the ambient acoustic component. For all the above cases we compute the Source to Interference Ratio (SIR) metric. The results obtained clearly demonstrate an optimum close-miking performance that matches the current empirical knowledge of professional audio recording.
\end{abstract}
}

\begin{document}

\maketitle

\section{Introduction}
Capturing sound through electro-acoustic transducers is one of the fundamental tasks in audio engineering. In practice, although audio recording is not restrained by particular specifications, there are applications where certain restrictions apply, for example when capturing an audio source output in the presence of other active audio sources (e.g. at a live music performance, or in a live recording session). In these cases, the presence of the latter sound sources introduces ambient noise, which is added to the ambient noise of the recording space (if any).

A widely-employed technique that is used for capturing an audio source that is simultaneously active with other sources is commonly known as close miking~\cite{huber2005}. It defines the microphone's placement close to the sound source and nearly in all cases it is used as a rule of thumb~\cite{kokkinis-12}. With this specific placement of the microphone, the captured audio tends to contain more energy from the targeted source than from the surrounding ones. Hence, close miking effectively functions as a mechanical source separation method that aims to separate the signal of the targeted source from the mixture of the sound field that is created by all concurrently active sound sources. The suggested distance for the microphone placement roughly spans from $0.03$ to $1$ meter away from the targeted source~\cite{huber2005}, balancing the trade off between affecting the timbre of the targeted source and the pickup of unwanted sources. Although this technique is widely--used, according to authors' best of knowledge, there is no previous study that systematically verified the above microphone distance or evaluated its effect on the resulting captured audio in terms of source separation.  

The field of source separation is not recent. It regards the estimation of individual signal components, denoted as sources, from their observed mixtures, and 
there are numerous published works focusing on this paradigm~\cite{liu2009}. Source separation has been utilized in many applications spanning from audio signal processing, e.g. for audio up-mixing\cite{fitzgerald2010}, stereo image enhancement~\cite{floros11, drosos2012}, harmonic-percussive separation \cite{cano}, source modeling \cite{kam} and singing voice/solo separation \cite{rafii_repet, mim16_dnn},
to neurological studies, for separating different electrical sources during physiological signals measurements~\cite{choi2005}, and satellite images, e.g. for detecting the actual morphology of the ground~\cite{loghmari2006}.

For evaluating source separation techniques, a couple of strategies have been proposed. More specifically, in ~\cite{bss_eval} a set of metrics are presented that can assess the extracted information from the mixture taking into account the produced artifacts (i.e. deformations induced by the separation algorithm, such as musical noise), noise (energy perturbations that does not correspond to the extracted source nor the interfering ones) and interference (a deformation of unwanted sources contributing to extracted information). Focusing on modeling and measuring the interference of unwanted sources subject to a targeted one, the notion of disjointness orthogonality is introduced in \cite{yilmaz_masks}. Assuming that non--interfering sources are completely orthogonal to each other in a signal domain, i.e. the short-time Fourier transform (STFT), the degree of overlap that the sources might have can be estimated providing an intuitive estimation of the total interference \cite{yilmaz_masks}.

Since close miking aims at separating the targeted source from the mixture of the total sound field that is created by all the active sound sources, it can be considered as a source separation technique and its effect to realistic scenarios
can be evaluated by the above mentioned strategies.
In this work we try to evaluate the close miking technique under the above perspective. We employ the aforementioned method for source separation evaluation based on the orthogonality assumption \cite{yilmaz_masks}, and assess the effect of distance, targeted sound source sound pressure level, interfering noise sound pressure level, angle of the microphone with respect to the central axis of the targeted sound source, and different types of microphones lobes by means of \textit{signal to interference ratio} (SIR), essentially objectifying the choice of microphone placement. For that cause, we conducted a series of measurements in a reverberant room, i.e. an empty theater, with two sound sources and a sound level meter for calibrating the reproduction levels.

The rest of the paper is organized as follows. Section~\ref{sec:overview} provides an overview of the existing literature that focuses on close miking, along with the presentation of the appropriate metrics and their computation. Section~\ref{sec:procedure} outlines the methodology followed for the performed measurements, while Section~\ref{sec:results} contains the obtained results. Finally, Section~\ref{sec:discussion} holds the discussion of the results and Section~\ref{sec:conclusions} concludes the paper and proposes future works.

\section{Existing work}\label{sec:overview}
\vspace{-7pt}
\subsection{Close miking technique}
\vspace{-5pt}
Close miking is rather based on empirical knowledge and a set of general guidelines that define the location and distance of the microphone from the sound source~\cite{huber2005}. Existing studies are particularly focusing on two different aspects. The first considers the varying spectral information and perceived timbre of the music sound sources. The second regards the inspection of the close miking technique from a signal processing point of view and its relation to room acoustics. 

Focusing on the first aspect, in~\cite{bartlett-2002} recordings of a variety of musical instruments and human voice are employed. These recordings are performed using different microphone placement distances, ranging from $0.03$ to $1$ meter. The recorded signals are transformed into the frequency domain and compared with the emanation patterns of each sound examined source. As an outcome, different equalization techniques are proposed depending on the placement of the recording microphone. Following the same approach, a work more centered to human voice is presented in~\cite{brixen-98}. It examines the distance of the placement of the microphone and its effect on the perceptual spectral content. Finally, in~\cite{case-10} an assessment of microphone placement with respect to the ambience reflections, transmitted to the recording device, and timbre is presented. Different microphone--source distances are examined alongside various angle orientations of the microphone with respect to the central axis of the sound source (i.e. $[15^{\circ},30^{\circ},45^{\circ},60^{\circ},90^{\circ}]$). 

Differentiating from the above studies, the work in \cite{kokkinis-12} evaluates close miking from a different signal processing perspective. In particular, this work aims to validate the close miking technique by examining the effect of the excitation of the surrounding acoustic space. To do so, sound sources are recorded in various distances and the recorded signals are subjectively assessed for their perceptual suppression of the reverberation effect. Nonetheless, all the literature described above relies on the empirical knowledge of the relative distance between the microphone and the sound source. A quantified answer regarding the definition of this distance range is still not being proposed.

\subsection{Computation of SIR}\label{sec:sig_mod}
For the evaluation of the source separation capabilities of the close miking technique, we employed the Signal to Interference Ratio (SIR) metric. Usually, this metric is used in the evaluation of the source separation task and indicates the energy ratio between a signal, separated from mixture of signals, and the interference from the mixture that is apparent in the separated signal. 

More formally, let $\mathbf{x}$ be a vector denoting a single-channel (\textit{monaural}) mixture consisting of $2$ additive sources expressed as vectors $\mathbf{s}$ and $\mathbf{n}$.
Given that each source is known beforehand, the degree of overlap that the targeted source $\mathbf{s}$ and the interfering $\mathbf{n}$ have, can be computed yielding the objective measure of SIR.

To do so, an analysis operator $\Tau$ is applied to each source (targeted and interfering one) as follows:
\begin{align}
    \label{eq:tf_decomp}
	S(m, k) &= \Tau(\mathbf{s}),\\
	N(m, k) &= \Tau(\mathbf{n}),
\end{align}
where $\Tau$ corresponds to the STFT analysis operation using the parameters proposed by a standard source separation evaluation (SSE) scheme \cite{bss_eval-guide}, and $m$, $k$ denote the time-frames and frequency bins (sub-bands), respectively.

For the computation of $SIR$, given a pair of sources, the method presented in \cite{yilmaz_masks} is followed. Therefore, a time-frequency filtering operation, i.e. \textit{time-frequency masking}, is derived from Eq.~\ref{eq:tf_mask}:
\begin{equation}
\label{eq:tf_mask}
M(m, k) = \begin{cases}
1, \text{ if } {|S(m, k)|} \geq {|N(m, k)|} \\
0,  \text{ otherwise.}
\end{cases}
\end{equation}

Then, by taking into account all the available time-frequency samples $m$, $k$ and expressing as matrices the output of equations~\ref{eq:tf_decomp} --\ref{eq:tf_mask}, the $SIR$ is computed as follows:
\begin{equation}
	SIR = 10  \log_{10}\bigg(\frac{||\mathbf{M} \odot |\mathbf{S}| ||^2_{F}}{||\mathbf{M} \odot |\mathbf{N}|||^2_{F}}\bigg), 
\end{equation}
where $|\cdot|$ refers to the modulus, i.e. the \textit{magnitude}, of the time-frequency representation of each source, $\odot$ is an element-wise multiplication, and $||\cdot||^2_{F}$
denotes the squared Frobenious norm.

The values of $SIR$ will approach $+\infty$ when the magnitude of the acquired signal $S(m,k)$, for each time-frame and frequency sub-band, will be superior to the interfering one. On the other hand, when the values approach $-\infty$, then the interfering source completely dominates over their mixture. Essentially, this leads to a straightforward assessment of how well a method describes or estimates the targeted signal $\mathbf{x}$, in presence of outliers, can be acquired.

\section{Experimental procedure}\label{sec:procedure}
The experimental procedure of the work at hand is separated in two tasks: i) recording of the individual signals, and b) the computation of $SIR$ subject to each recording of a pair of sources. The former was utilized in a municipal theater, located in Lixouri, Kefalonia, (Ionian islands, Greece), before the disastrous earthquakes in the Autumn of $2014$ and resulted into the formation of the audio dataset employed by the SSE task. The latter was implemented by utilizing the signal model described in Section~\ref{sec:sig_mod}.

The aim of the first task is to provide a comprehensive set of recorded material containing the source signal, the noise signal, and the mixture of both. Each recorded waveform is characterized by: a) the distance between the microphone and the signal source, b) the type of microphone, and c) the sound pressure level (SPL) of the actual source and the noise source. Various combinations of the above factors were considered in the particular task. On the other hand, the second task involves the evaluation of the performance of close miking as a source separation method. The expected outcome is to determine the effective limits and the relations between the key factors mentioned above, subject to an objective measure. In the following sections the above tasks will be presented in detail. 

\begin{table}
	\centering
	\caption{List of the equipment used for audio recordings}
	\label{tab:equip}
	\scalebox{.9}{
		\begin{tabular}{m{1.6cm} m{1.8cm} | m{1.6cm} m{1.7cm} }
			\textbf{Apparatus} & \textbf{Model} & \textbf{Apparatus} & \textbf{Model}\\
			\hline
			SLM & \small B\&K $2250$ Type A SLM & Mic. A & \small Shure SM57, dynamic, cardioid\tabularnewline
			Laptop & \small Macbook Pro 15'' & Mic. B & \small Behringer ECM8000, condenser, omni-directional\tabularnewline
			\small Recording software & \small Digidesign ProTools M-Powered 8& \small Musical instrument amplifier  & \small Behringer V-Tone GMX212\tabularnewline
			\small Digital sound card & \small M-Audio Fast Track Ultra& \small{Loudspeaker} & \small Electrovoice SX300 \tabularnewline
		\end{tabular}
	}
\end{table}

\subsection{Recordings procedure}
The audio recordings were performed using a musical instrument amplifier, one loudspeaker, one laptop with recording software and a digital sound card, two microphones (one dynamic and one condenser/measurements), and one Sound Level Meter (SLM). One microphone was omni-directional, while the the other had a cardioid lobe. The full list of all equipment parts is provided in Table~\ref{tab:equip}.

Close miking aims at diminishing the addition of the noise in the final audio mixture. The prime element that affects the efficacy of this technique is the distance between the microphone and the sound source. But since, on one hand, the distance between the microphone and the actual sound source can result into an attenuation of the SPL and, on the other hand, different sound sources in a real-world scenario are likely to exhibit varying SPL, the question of the effect of SPL in the close miking technique is also raised. Finally, various receiving patterns of microphones are utilized in a recording session. These affect the effective SPL recorded by the microphone and thus different microphone lobes are possible to portray divergent results in close miking. In addition, there are references in the utilization of an angle between the central axes of the microphone and the sound source in order to achieve improved attenuation of the receiving noise from the microphone. 

\begin{figure}[!h]
\centering
\setlength{\unitlength}{3947sp}%
\begingroup\makeatletter\ifx\SetFigFont\undefined%
\gdef\SetFigFont#1#2#3#4#5{%
  \reset@font\fontsize{#1}{#2pt}%
  \fontfamily{#3}\fontseries{#4}\fontshape{#5}%
  \selectfont}%
\fi\endgroup%
\begin{picture}(2300,2307)(7051,-944)
\put(7931,-771){\makebox(0,0)[lb]{\smash{{\SetFigFont{9}{10.8}{\rmdefault}{\mddefault}{\updefault}{\color[rgb]{0,0,0}Card}%
}}}}
{\color[rgb]{0,0,0}\thinlines
\put(7924,-456){\circle{324}}
}%
{\color[rgb]{0,0,0}\put(7328,-284){\line( 1, 0){807}}
\put(8135,-284){\line( 0, 1){  6}}
}%
{\color[rgb]{0,0,0}\multiput(7719,153)(0.00000,125.20000){3}{\line( 0, 1){ 62.600}}
\put(7719,153){\vector( 0,-1){0}}
\multiput(7724,466)(125.09091,0.00000){6}{\line( 1, 0){ 62.545}}
}%
{\color[rgb]{0,0,0}\multiput(8780,-32)(0.00000,122.57143){4}{\line( 0, 1){ 61.286}}
\put(8780,-32){\vector( 0,-1){0}}
}%
{\color[rgb]{0,0,0}\multiput(8780,622)(0.00000,123.11111){5}{\line( 0, 1){ 61.556}}
\put(8780,1176){\vector( 0, 1){0}}
\put(8780,622){\vector( 0,-1){0}}
}%
{\color[rgb]{0,0,0}\multiput(7759,-502)(0.00000,-122.85714){4}{\line( 0,-1){ 61.429}}
\multiput(7759,-932)(-126.54545,0.00000){6}{\line(-1, 0){ 63.273}}
\multiput(7063,-932)(0.00000,119.84000){13}{\line( 0, 1){ 59.920}}
\multiput(7063,566)(117.65217,0.00000){12}{\line( 1, 0){ 58.826}}
\put(8416,566){\vector( 1, 0){0}}
}%
{\color[rgb]{0,0,0}\put(8656,-850){\framebox(227,812){}}
}%
{\color[rgb]{0,0,0}\put(7333,-33){\framebox(835,195){}}
}%
{\color[rgb]{0,0,0}\put(8416,431){\framebox(697,197){}}
}%
{\color[rgb]{0,0,0}\put(8212,1175){\framebox(1127,176){}}
}%
\put(8818,-838){\rotatebox{90.0}{\makebox(0,0)[lb]{\smash{{\SetFigFont{9}{10.8}{\rmdefault}{\mddefault}{\updefault}{\color[rgb]{0,0,0}Noise}%
}}}}}
\put(7349, 17){\makebox(0,0)[lb]{\smash{{\SetFigFont{9}{10.8}{\rmdefault}{\mddefault}{\updefault}{\color[rgb]{0,0,0}Source}%
}}}}
\put(8476,487){\makebox(0,0)[lb]{\smash{{\SetFigFont{7.5}{10.8}{\rmdefault}{\mddefault}{\updefault}{\color[rgb]{0,0,0}Audio I/O}%
}}}}
\put(8244,1223){\makebox(0,0)[lb]{\smash{{\SetFigFont{8}{10.8}{\rmdefault}{\mddefault}{\updefault}{\color[rgb]{0,0,0}Recording Device}%
}}}}
\put(7286,-762){\makebox(0,0)[lb]{\smash{{\SetFigFont{9}{10.8}{\rmdefault}{\mddefault}{\updefault}{\color[rgb]{0,0,0}Omni}%
}}}}
{\color[rgb]{0,0,0}\put(7598,-454){\circle{324}}
}%
\end{picture}%
\caption{The set-up of the measurements}
\label{img:exp_rig}
\end{figure}
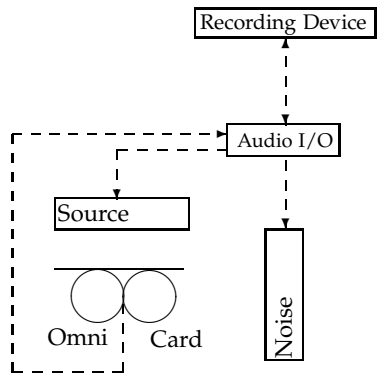

\begin{figure}
	\centering
	\includegraphics[width=.9\columnwidth]{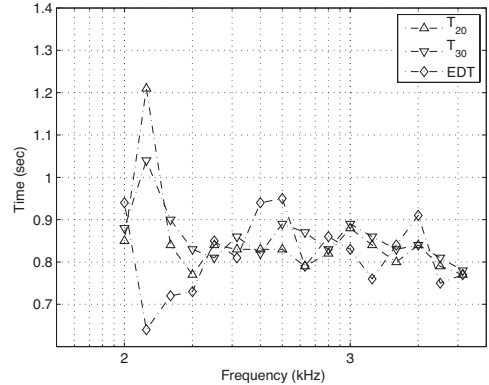}
	\caption{$T_{20}$, $T_{30}$ and $EDT$ measured in the stage of the theater at which the recordings took place}
	\label{img:reverberation_measures}
\end{figure}

In order to allow the investigation of the source--microphone distance, the source's SPL, microphone's lobe and microphone-sound source angle's effect in the particular technique, the experimental set-up presented in Figure~\ref{img:exp_rig} was performed: $2$ audio sources, $1$ laptop, $1$ digital sound card and two microphones were utilized for the recording. For the angle case, the cardioid microphone was used. The details of each component are listed in Table~\ref{tab:equip}. Thus, the loudspeaker served as the noise source, the musical instrument amplifier as the targeted sound source and the other are self--explanatory with respect to their utilization in the experimental process. All apparatuses employed are rather common to music performances, a case where close miking technique is thoroughly met. As source signal, a clean guitar riff (repetitive tonal sound from electric guitar, amplified through the corresponding amplifier) was employed without applying any distorting sound effects. In order to introduce ambient noise, a pink noise generator was activated. Each signal had a time length of $15$ seconds. 

The recording process consisted of two phases. The first realized the reverberation measurement of the recording room, while the second included the actual recordings. Regarding the latter case, different recordings were considered with a) the omni-directional lobe microphone, b) the cardioid microphone with its central axis aligned with the central axis of the sound source and c) the cardioid microphone placed with an angle of $45$ degrees relative to the sound source's central axis. The reverberation measurement was implemented with the use of the SLM at six different positions in the stage of the theater, forming a hexagon. For all positions the $T_{20}$, $T_{30}$ and Early Decay Time ($EDT$) values were obtained. The results are illustrated in Figure~\ref{img:reverberation_measures}.

\begin{table}
	\centering
	\caption{Sound source - microphone distances used in the first phase of the recording procedure}
	\label{tab:experimental_procedure_distances}
	\begin{tabular}{c c | c c}
		\textbf{Index $\mathbf{i_d}$} & \textbf{Distance (m)} & \textbf{Index} & \textbf{Distance (m)}\tabularnewline
		\hline
		01 & 0.03 & 07 & 0.21\tabularnewline
		02 & 0.06 & 08 & 0.24\tabularnewline
		03 & 0.09 & 09 & 0.27\tabularnewline
		04 & 0.12 & 10 & 0.30\tabularnewline
		05 & 0.15 & 11 & 0.65\tabularnewline
		06 & 0.18 & 12 & 1.00\tabularnewline
	\end{tabular}
\end{table}

Clearly, the theater stage can be considered fairly reverberant, especially in the region of $2.2kHz$. This fact allows our investigation to be performed in a rather un-favored environment; thus it can provide results that correspond to cases where close miking would be favored in order to eliminate capturing of audio signals emerging from all noise sources, including ambient noise. Regarding the second phase of the recordings process, $12$ different sound source-microphone distances were employed, with index $i_d\in[1,12]$ and ranging from $0.03$ to $1$ meter. The third recording type (with  the cardioid microphone placed with an angle of $45$ degrees relative to the sound source's central axis) included $10$ additional distances, ranging from $0.03$ to $0.30$ meters, marked as $i_d\in[1,10]$. For clarity, these values are summarzied in Table~\ref{tab:experimental_procedure_distances}. Up to $0.3$ meters the distance increment step equals to $0.03$ meters. Above that limit, it becomes $0.35$ meters. The reason for that is the apparent evidence in the existing literature, that above $0.3$ meters close miking technique suffers from leakage and interference, when the sound pressure level of unwanted sources is high, contrary to the desired source \cite{huber2005, kokkinis-12}. Also, in the second phase the utilized distances are those with index $i_d\leq10$. Moreover, different SPL values were employed for both sound and noise sources and for all microphone lobe's cases. For the former sound source, a set of $2$ different SPL, $SPL_S[i_S],\,i_S\in[1,\,2]$, values were used, whereas for the latter a set of $5$, $SPL_N[i_N],\,i_N\in[1,5]$. This information is listed in Table~\ref{tab:experimental_procedure_spl}. 

\begin{table}
	\centering
	\caption{SPL values used for the recordings procedure}
	\label{tab:experimental_procedure_spl}
	\begin{tabular}{cc}
		\textbf{Index} & \textbf{SPL, ref $\mathbf{P_{ref}=2\times10^{-4}}$}\tabularnewline
		\hline
		\multicolumn{2}{c}{\textbf{Sound Source SPL ($\mathbf{SPL_{S}}$)}}\tabularnewline
		\hline
		$SPL_{S}[1]$ & $100$ dB SPL\tabularnewline
		$SPL_{S}[2]$ & $97$ dB SPL\tabularnewline
		$SPL_{S}[3]$ & $94$ dB SPL\tabularnewline
		\hline
		\multicolumn{2}{c}{\textbf{Noise Source SPL ($\mathbf{SPL_{N}}$)}}\tabularnewline
		\hline
		$SPL_{N}[1]$ & $100$ dB SPL\tabularnewline
		$SPL_{N}[2]$ & $97$ dB SPL\tabularnewline
		$SPL_{N}[3]$ & $94$ dB SPL\tabularnewline
		$SPL_{N}[4]$ & $91$ dB SPL\tabularnewline
		$SPL_{N}[5]$ & $88$ dB SPL\tabularnewline
	\end{tabular}
\end{table}

The different SPLs for the sound and noise source have a variation step of $3$ dB SPL, since this difference corresponds to two times the acoustic energy. Also, there are $3$ different $SPL_S$: one that can be considered as high, one  as medium and one as low. In conjunction with the $SPL_{N}$, these values allow the investigation of the different SPL effect. More specifically, each $SPL_{S}$ was used with every $SPL_{N}$, i.e. for $SPL_{S}$ all $SPL_{N}$ were utilized for the noise source and the same stands for $SPL_{S}[2]$. Thus, for $SPL_{S}[1]$ it can be seen that the selected $SPL_{N}$ span in the dynamic range of equal SPL to $1/24$ times lower (for the case of $SPL_{S}[1]$ and $SPL_{N}[5]$). In the case of $SPL_{S}[2]$, the dynamic range of SPL corresponds to double acoustic energy emerging from the noise source as well as the same, half, one quarter and one eighth acoustic energy for the noise. Regarding the $SPL_{S}[3]$ it can be seen that the selected SPL for the noise source corresponds to quadruple, double, equal, half and one quarter acoustic energy emerging from the noise source. All SPLs were calculated in terms of $L_{eq}$, with a time length average equal to the time length of both the sound and noise source signal (i.e. $15$ seconds). 

The actual recordings were performed for each microphone and for all $SPL_{S}$, $SPL_{N}$ and appropriate sound source-microphone distances. In particular, if $D[i_d]$ are the different distances as presented in Table~\ref{tab:experimental_procedure_distances}, $M_{t},t\in[1,2]$ are the different microphone lobes with $M_1$ to be the omni-directional and $M_2$ to be the cardioid lobe, $Ang[i_{ang}],\,i_{ang}\in[1,3]$ the angle between the microphone's and source's central axis, with $Ang[1]=0^{\text{o}}$, $Ang[2]=30^{\text{o}}$ and $Ang[3]=45^{\text{o}}$, since the effect is minimal  for lower angle variations \cite{case-10},  and $SPL_{S}[i_S]$ and $SPL_{N}[i_N]$ the different SPLs for the sound and noise source respectively, then the following recording sets, $R_{i}$, were created:

\begin{align}
	R_{1}=&\{D[i_d],\,M_{1},\,Ang[1],\,SPL_{N}[i_N]\}\label{eq:recording_set_1}\\
	R_{2}=&\{D[i_d],\,M_{1},\,Ang[1],\,SPL_{S}[i_S]\}\label{eq:recording_set_2}\\
	R_{3}=&\{D[i_d],\,M_{2},\,Ang[1],\,SPL_{N}[i_N]\}\label{eq:recording_set_3}\\
	R_{4}=&\{D[i_d],\,M_{2},\,Ang[1],\,SPL_{S}[i_S]\}\label{eq:recording_set_4}\\
	R_{5}=&\{D[i'_d],\,M_{2},\,Ang[2],\,SPL_{N}[i_N]\}\label{eq:recording_set_5}\\
	R_{6}=&\{D[i'_d],\,M_{2},\,Ang[2],\,SPL_{S}[i_S]\}\label{eq:recording_set_6}\\
	R_{7}=&\{D[i'_d],\,M_{2},\,Ang[3],\,SPL_{N}[i_N]\}\label{eq:recording_set_7}\\
	R_{8}=&\{D[i'_d],\,M_{2},\,Ang[3],\,SPL_{S}[i_S]\}\label{eq:recording_set_8}\\
	\nonumber
\end{align}

where $i_d\in[1,12]$, $i'_d\in[1,5]$, $i_S\in[1,3]$ and $i_N\in[1,5]$. It must be noted that in the cases where a recording contains both $SPL_S$ and $SPL_N$, these two were physically apparent and recorded at the same time. The calibration of the SPL for each sound source ($SPL_S$ and $SPL_N$) was performed with the SLM, at the point of the recording microphone, for each source-microphone distance separately, and without any other source active.

The recordings in the overall data set were all time trimmed to $15$ seconds in order to contain exactly the produced signals from all cases. The audio data from the $15$ seconds long recordings were saved under standard CD quality, i.e.sampling frequency equal to $44.1kHz$ and $16$ bit sample length, using the typical wave file format. The latter audio files were utilized by the SSE process presented immediately next, organized in the sets $R'_1$ to $R'_8$, in accordance to Equations~\ref{eq:recording_set_1} to~\ref{eq:recording_set_8}.

\begin{figure*}[!ht]
	\centering
	\begin{subfigure}{.33\textwidth}
		\includegraphics[width=\linewidth]{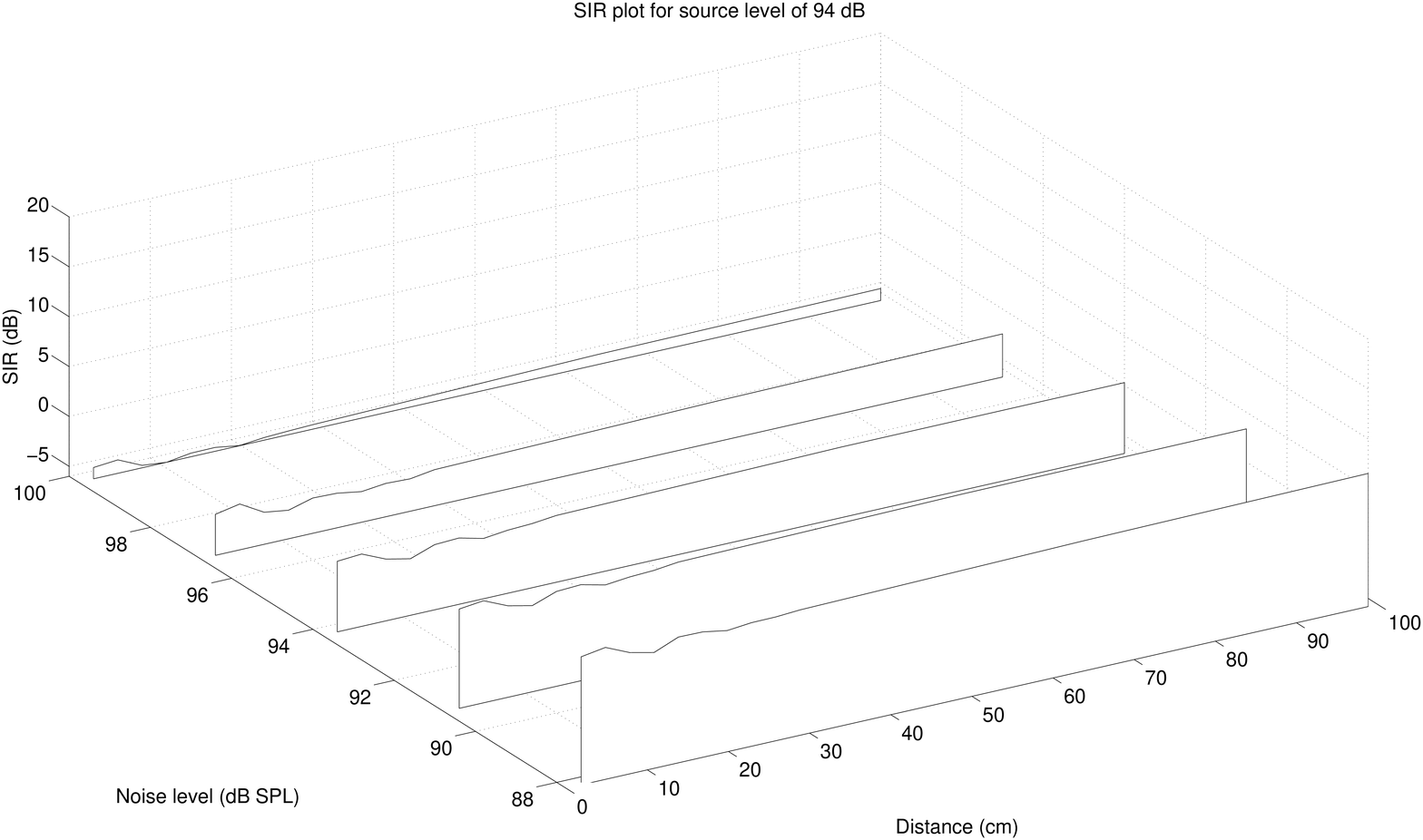}
		\caption{Source SPL : $94$ dB}
	\end{subfigure}
	\begin{subfigure}{.33\textwidth}
		\includegraphics[width=\linewidth]{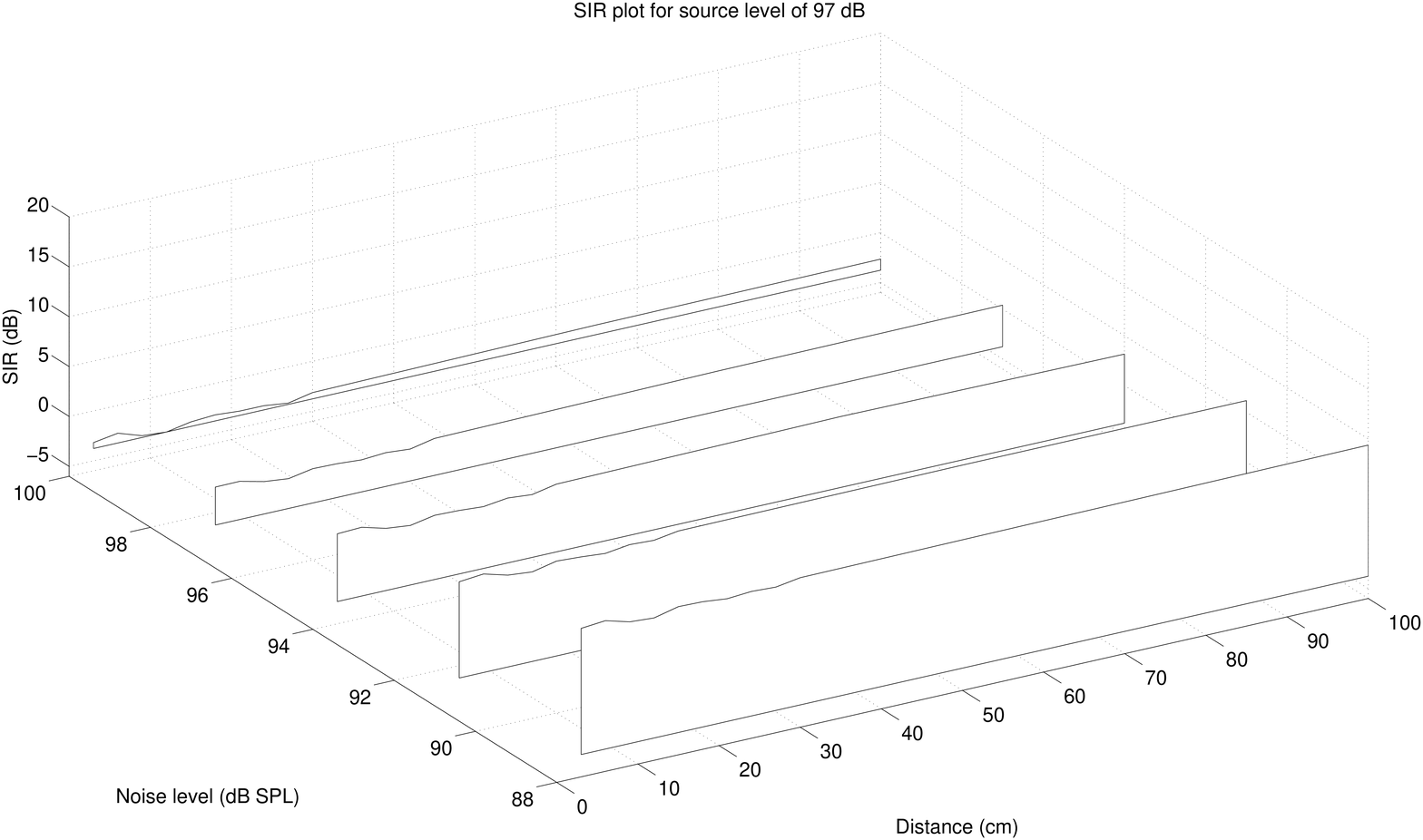}
		\caption{Source SPL : $97$ dB}
	\end{subfigure}
	\begin{subfigure}{.33\textwidth}
		\includegraphics[width=\linewidth]{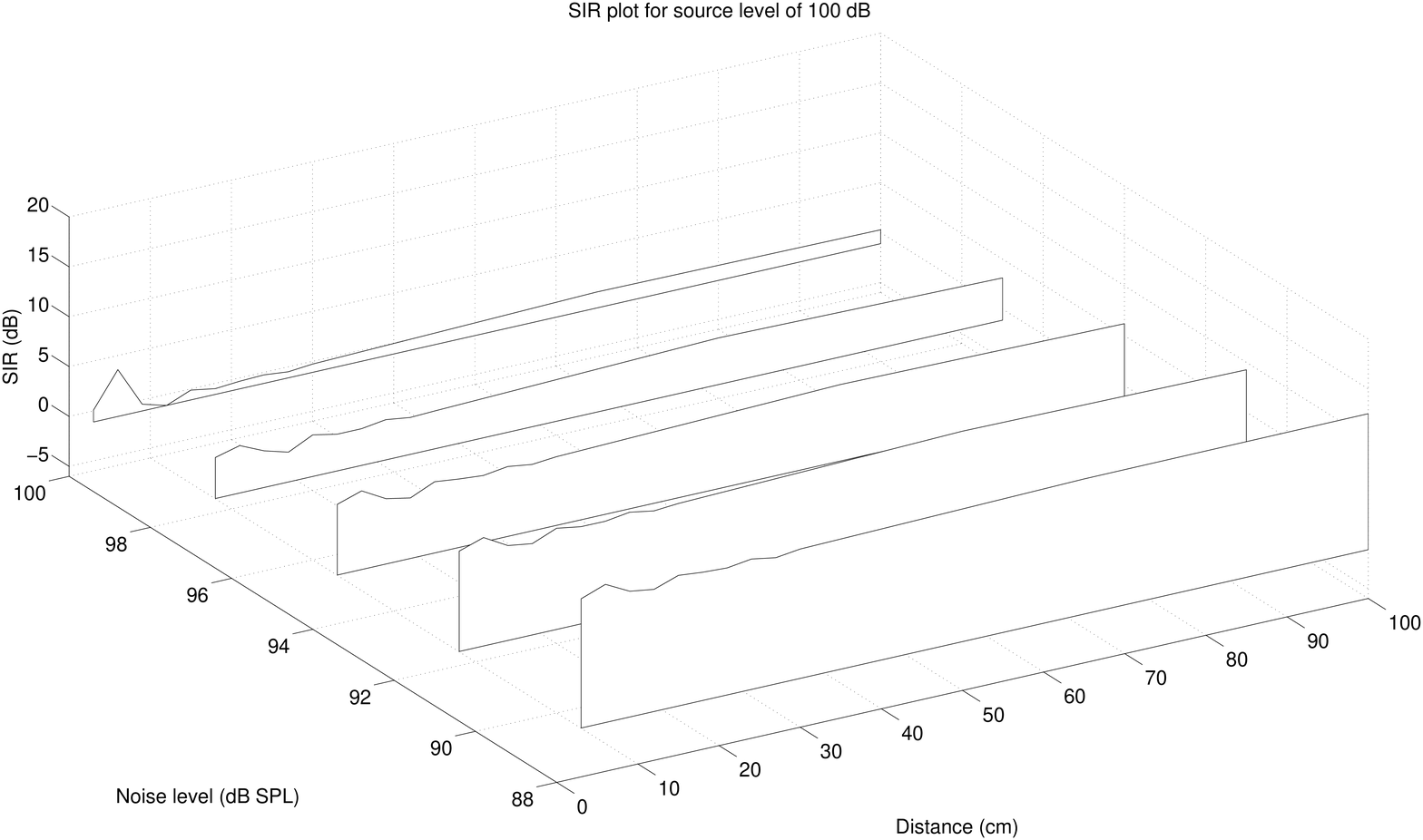}
		\caption{Source SPL : $100$ dB}
	\end{subfigure}
	\caption{SIR of omni directional microphone over various sound pressure levels of \textit{source},with respect to distance $D$ and sound pressure level of \textit{noise}.}
	\label{fig:omni_sirSSPL}
\end{figure*}

\subsection{Source separation evaluation}
For evaluation purposes, pairs of audio files from the recording sets were utilized as input to the SSE process. Each pair contains two audio files, one containing the noise-free recording (i.e. the desired source is active only; an audio file from recording sets with even index), considered as the estimated source in terms of the SSE process, and the audio file from the recording with the noise source active (i.e. audio file from recording sets with odd index).

\begin{figure*}
	\centering
	\begin{subfigure}{.33\textwidth}
		\includegraphics[width=\linewidth]{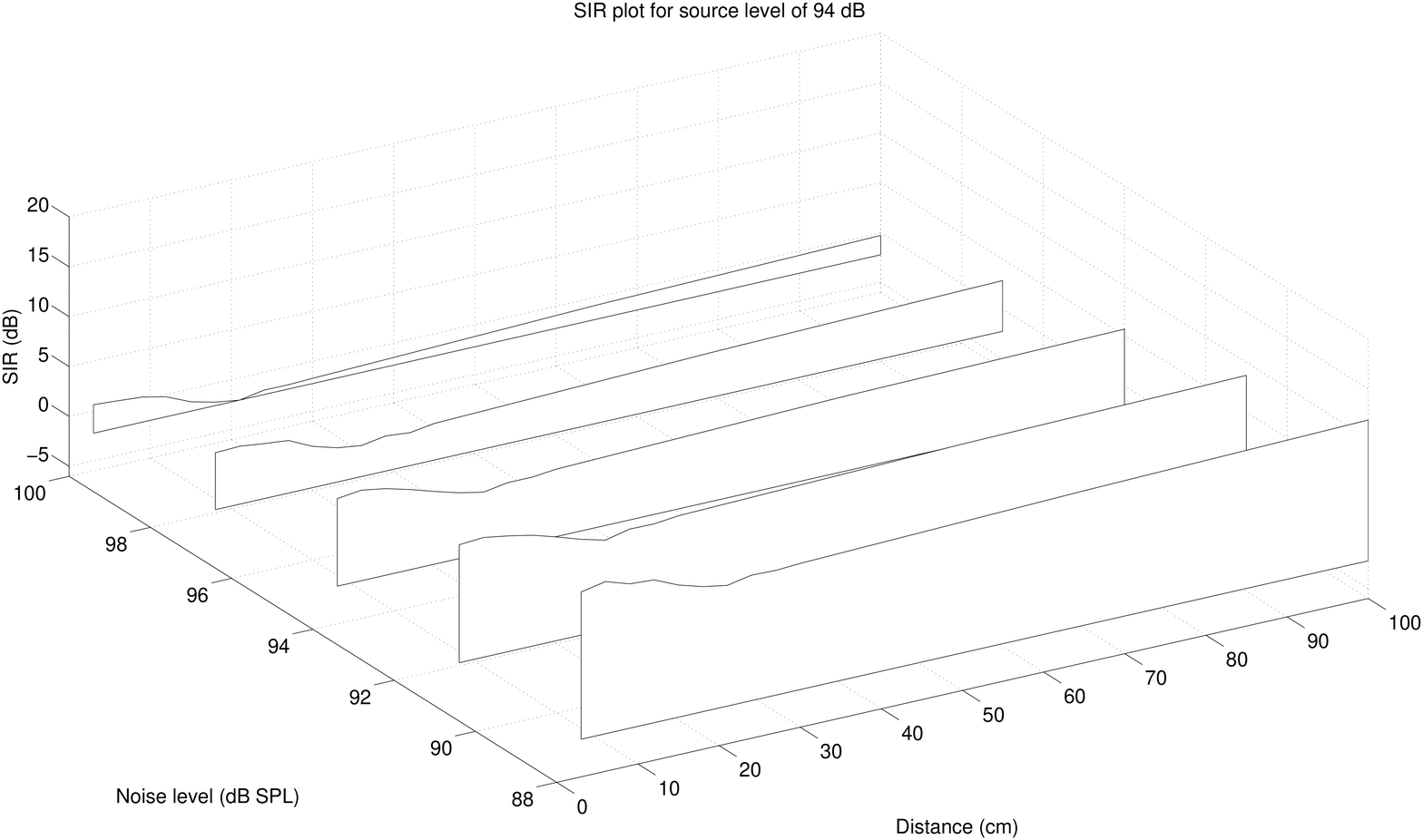}
		\caption{Source SPL : $94$ dB}
	\end{subfigure}
	\begin{subfigure}{.33\textwidth}
		\includegraphics[width=\linewidth]{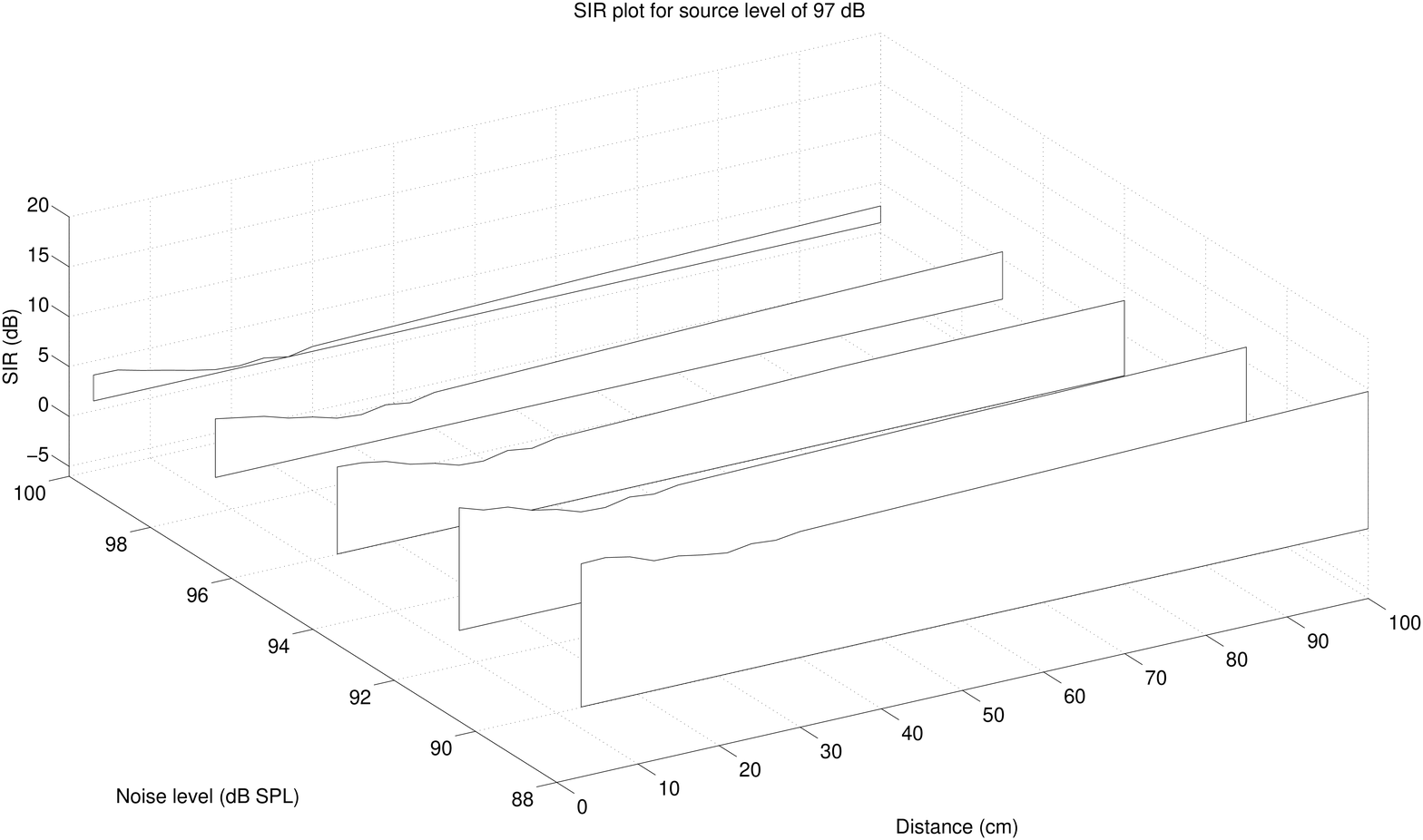}
		\caption{Source SPL : $97$ dB}
	\end{subfigure}
	\begin{subfigure}{.33\textwidth}
		\includegraphics[width=\linewidth]{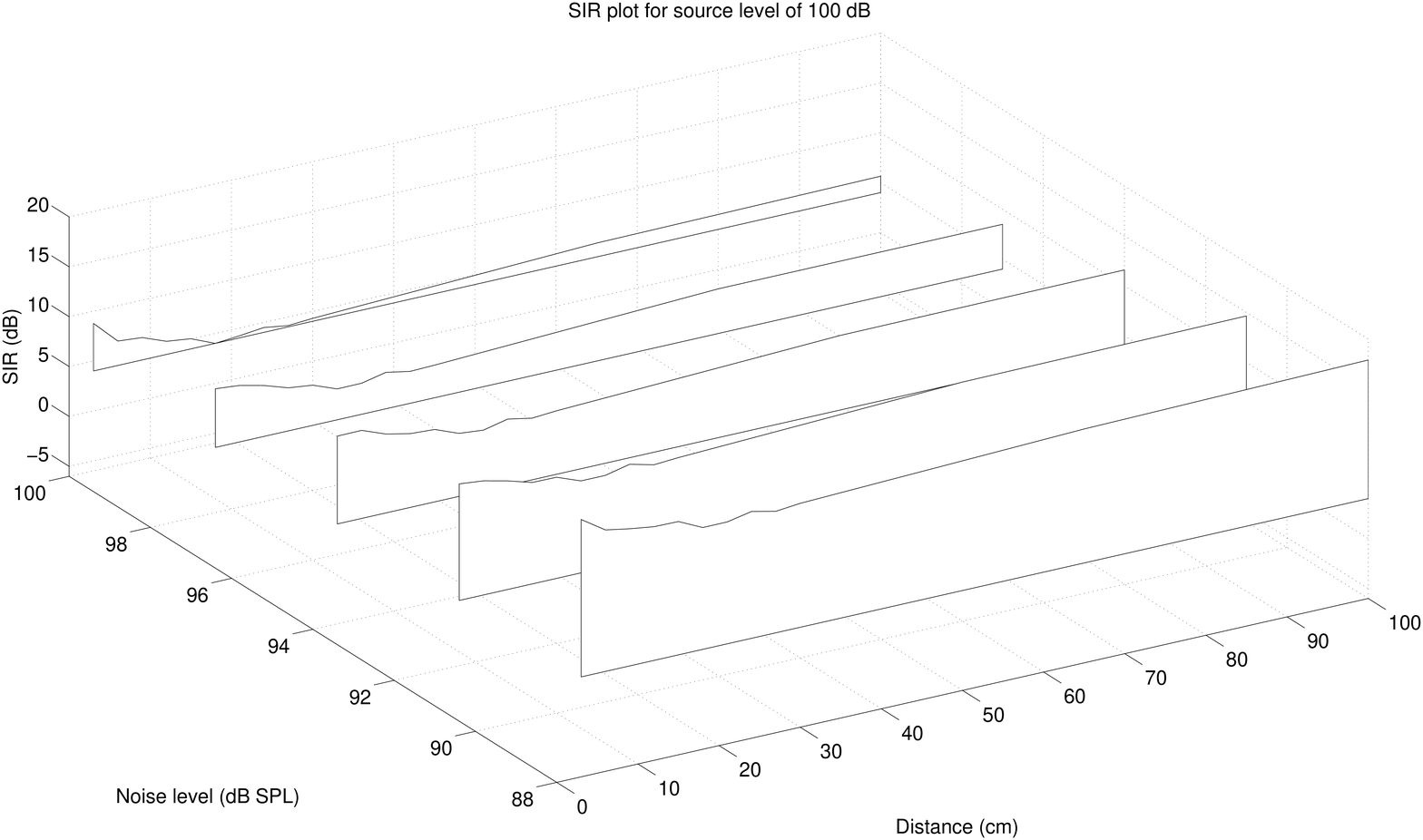}
		\caption{Source SPL : $100$ dB}
	\end{subfigure}
	\caption{SIR of cardioid microphone over various sound pressure levels of \textit{source}, with respect to distance $D$ and sound pressure level of \textit{noise}.}
	\label{fig:card_sirSSPL}
\end{figure*}

The SIR was computed for the recording set pairs: a) $R'_1$ and $R'_2$, b) $R'_3$ and $R'_4$, c) $R'_5$ and $R'_6$, and d) $R'_7$ and $R'_8$. As can be seen from Equations~\ref{eq:recording_set_1} to~\ref{eq:recording_set_8},  the recording sets with odd indices contain recordings with the noise source active and recording sets with even indices contain recordings with the desired source active. Also, each of the pairs a) to d), contains recording sets with the same microphone type and the same angle between the microphone and the sound source. Thus, the input for the calculation of the SIR for one recording pair was audio from each of the recording sets in this pair and with the same indices $i_d$/$i'_d$, $i_s$, $i_n$, and $i_{ang}$ 

\section{Results}\label{sec:results}
\begin{figure*}
	\centering
	\begin{subfigure}{.33\textwidth}
		\includegraphics[width=\linewidth]{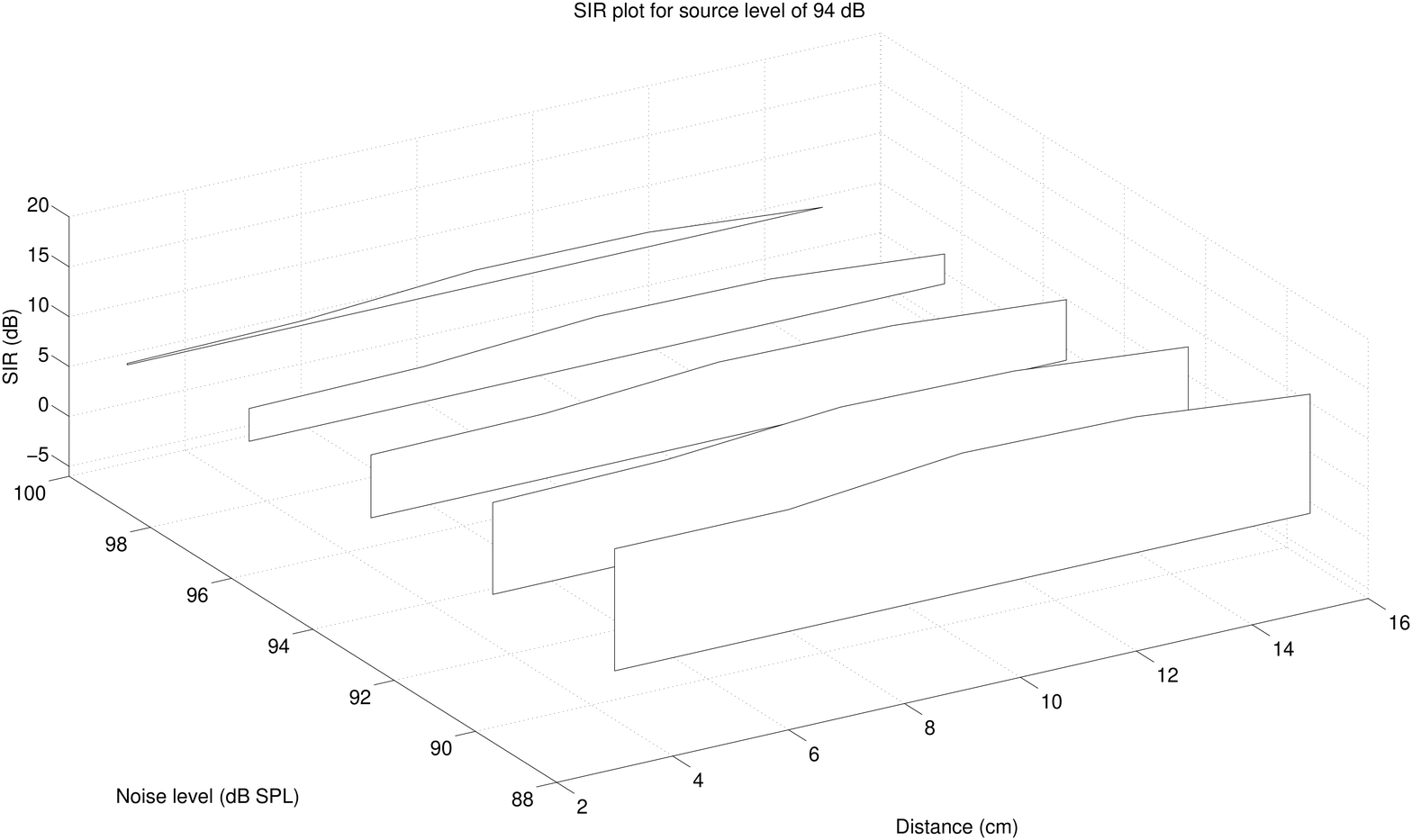}
		\caption{Source SPL : $94$ dB}
	\end{subfigure}
	\begin{subfigure}{.33\textwidth}
		\includegraphics[width=\linewidth]{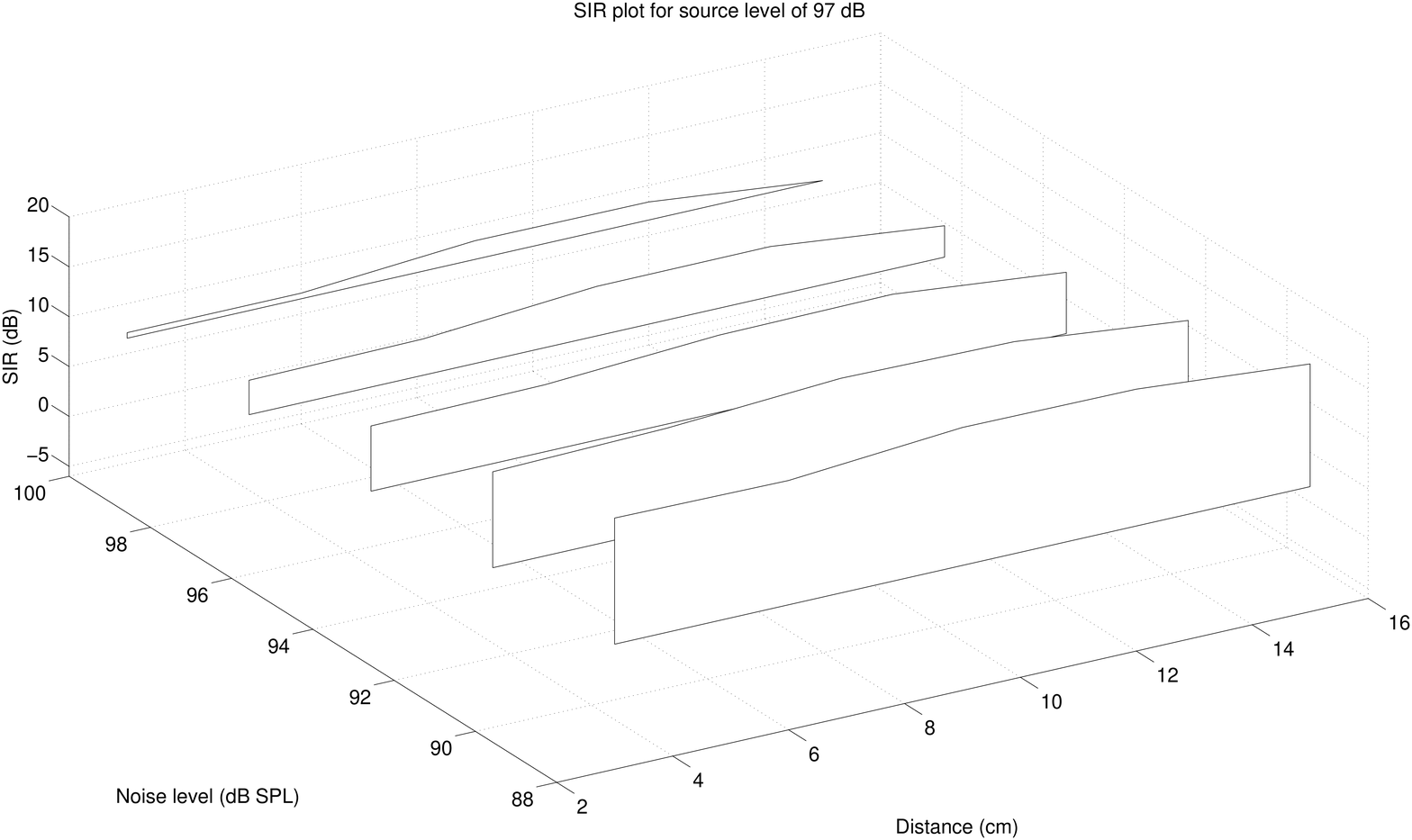}
		\caption{Source SPL : $97$ dB}
	\end{subfigure}
	\begin{subfigure}{.33\textwidth}
		\includegraphics[width=\linewidth]{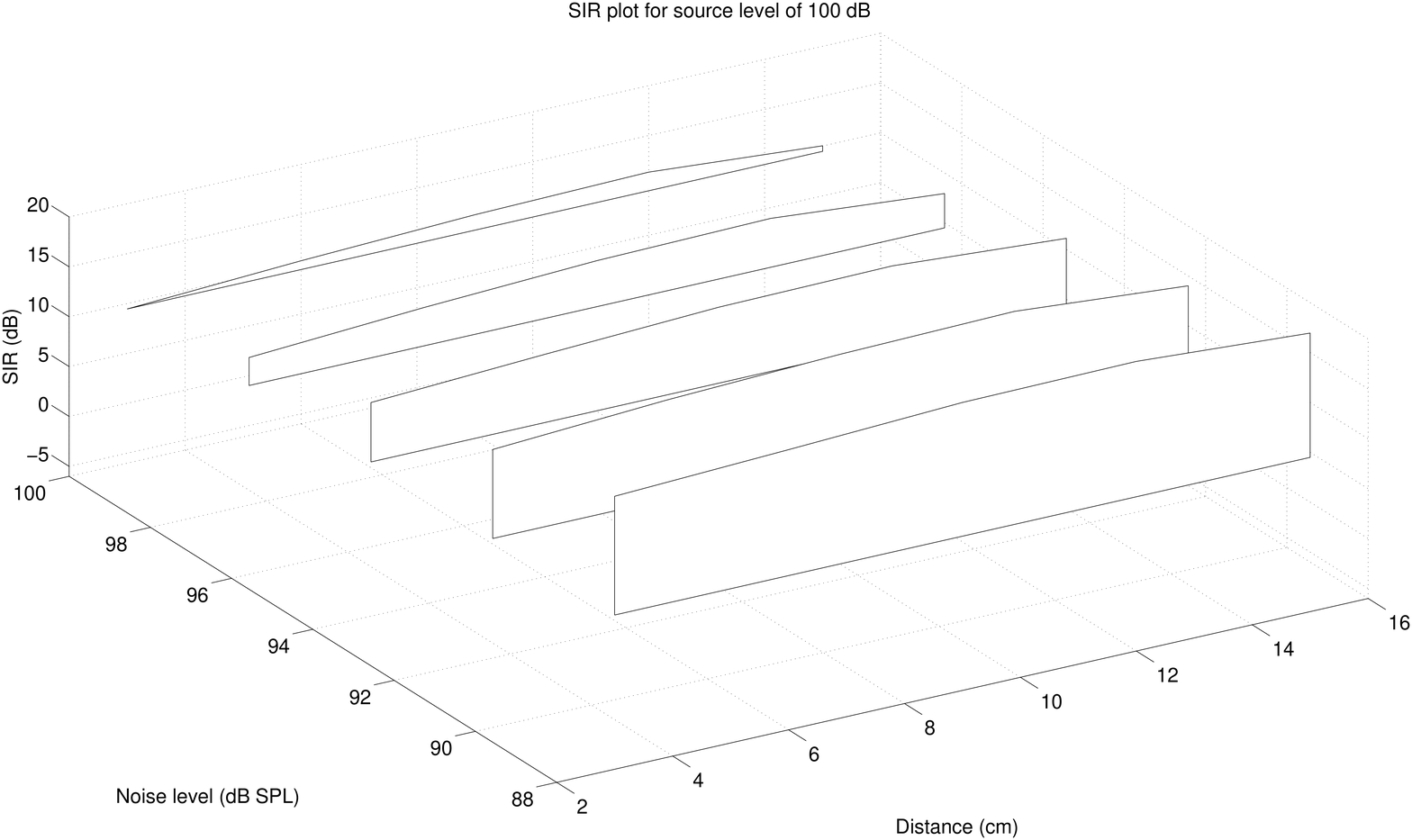}
		\caption{Source SPL : $100$ dB}
	\end{subfigure}
	\caption{SIR of cardioid microphone, with an angle of $30^\circ$, over various sound pressure levels of \textit{source}, with respect to distance $D$ and sound pressure level of \textit{noise}.}
	\label{fig:card_sirSSPL30}
\end{figure*}

\begin{figure*}
	\centering
	\begin{subfigure}{.33\textwidth}
		\includegraphics[width=\linewidth]{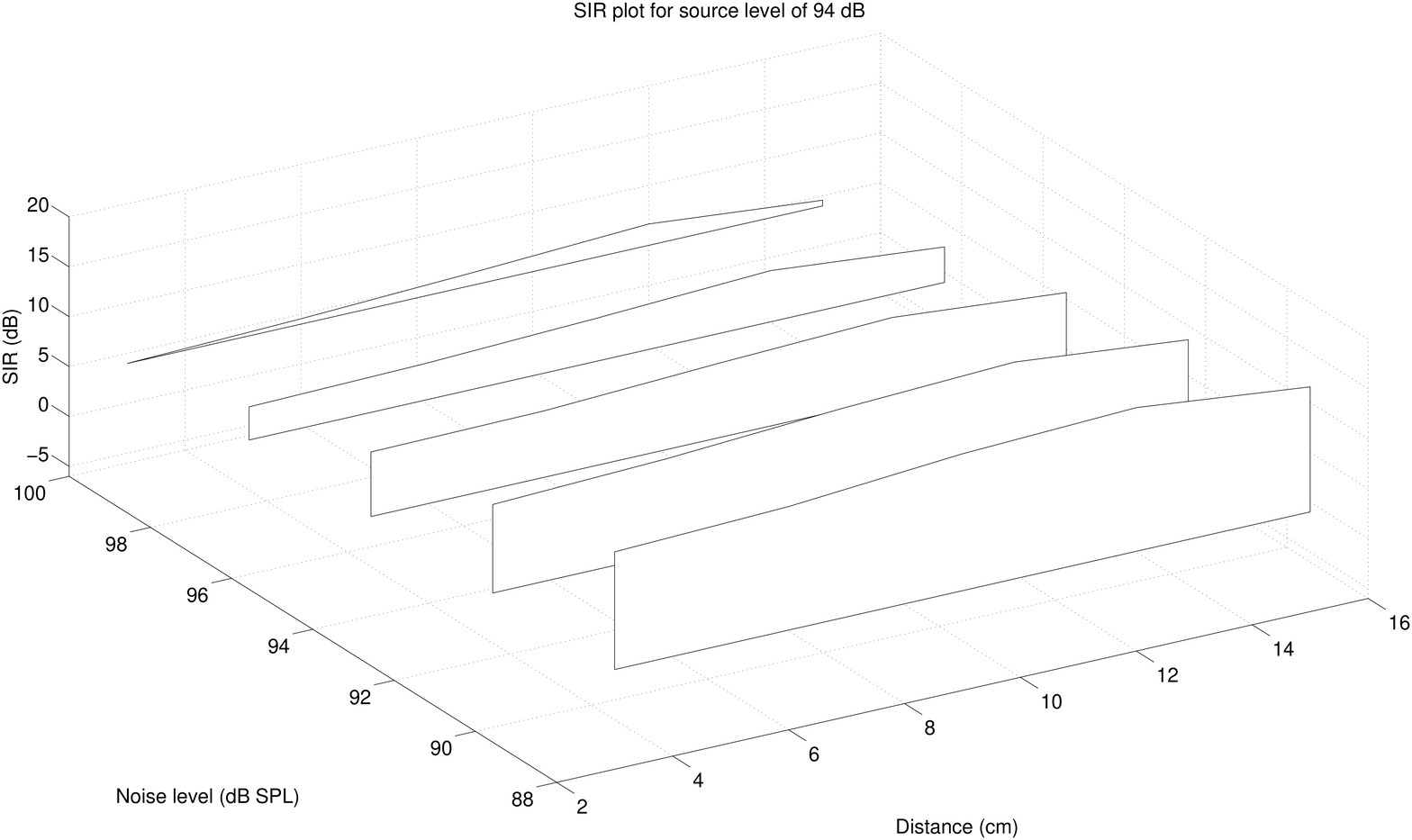}
		\caption{Source SPL : $94$ dB}
	\end{subfigure}
	\begin{subfigure}{.33\textwidth}
		\includegraphics[width=\linewidth]{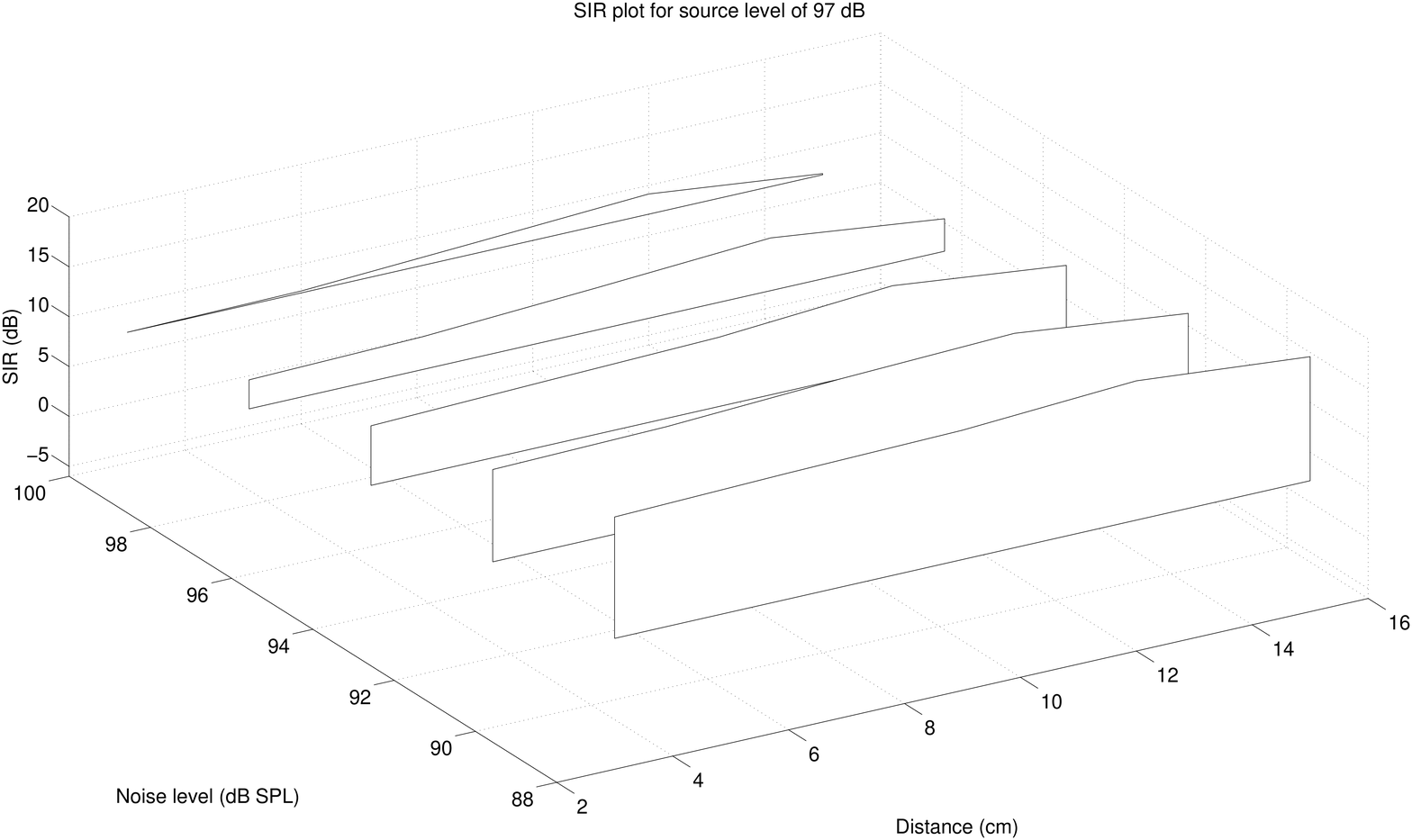}
		\caption{Source SPL : $97$ dB}
	\end{subfigure}
	\begin{subfigure}{.33\textwidth}
		\includegraphics[width=\linewidth]{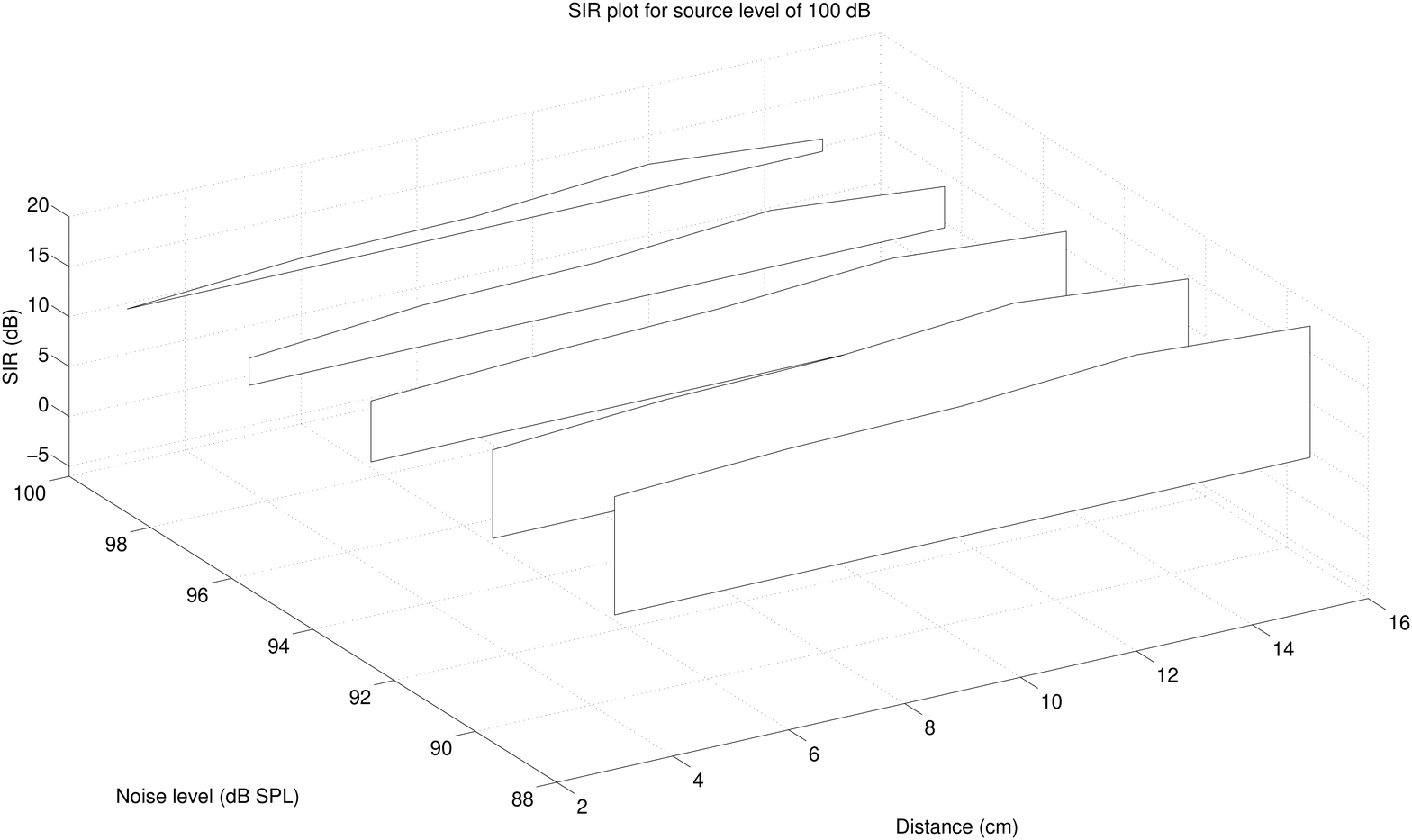}
		\caption{Source SPL : $100$ dB}
	\end{subfigure}
	\caption{SIR of cardioid microphone, with an angle of $45^\circ$, over various sound pressure levels of \textit{source}, with respect to distance $D$ and sound pressure level of \textit{noise}.}
	\label{fig:card_sirSSPL45}
\end{figure*}

\vspace{-8pt}
The results from the above experimental process are organized in $12$ figures, corresponding to the different combinations of microphone types, placement angles and the produced SPL. Specifically, in Figure~\ref{fig:card_sirSSPL} are the results for the cardioid microphone and for zero degrees angle between the microphone and the source. In Figure~\ref{fig:omni_sirSSPL} are the results for the omni-directional microphone. In Figure~\ref{fig:card_sirSSPL30} are the results for the cardioid microphone with an angle of $30^\circ$ between the microphone and the source and in Figure~\ref{fig:card_sirSSPL45} the results for the cardioid microphone and with an angle of $45^\circ$ between the microphone and the source are shown.

%

\section{Discussion}\label{sec:discussion}
The results presented in the previous section portray the expected fact that the lower SPL of the noise results in better performance of the close miking technique. Also, a general trend from all figures and subfigures is that the SPL of the source and the SIR seems to be analogous. This means that the higher the SPL of the source, the higher the SIR. These observations are in accordance with the general purpose and expectations of the close miking technique. 

Focusing on Figures~\ref{fig:omni_sirSSPL} and~\ref{fig:card_sirSSPL}, one can see that in all cases the cardioid microphone outperforms the omni-directional one. The SIR values obtained with the cardioid microphone are almost double of the SIR values obtained with the omni-directional. In addition, in both cases the maximum performance of the close miking technique seems to be achieved for a $5$ cm distance between the source and the microphone. After that distance, a reduction of the SIR is observed for both cases. For the omni--directional case, the reduction is between $10$ and $20$ centimeters (cm), while for the cardioid microphone case, the reduction is observed between $20$ and $40$ cm. Followed by that reduction, the SIR rises up to a limit achieved around $70$ cm. 

Focusing on Figures~\ref{fig:card_sirSSPL30} and~\ref{fig:card_sirSSPL45}, one can also observe better interference reduction (higher SIR values) for all source SPL, distances, and noise SPL when compared to the previous two cases. Additionally, in the same cases, i.e. Figures~\ref{fig:card_sirSSPL30} and~\ref{fig:card_sirSSPL45}, there is a maximum of SIR around $12$ to $14$ cm. This comes in contrast with the previous two cases where the peak was observed below $10$ cm. Since for the cases of Figures~\ref{fig:card_sirSSPL30} and~\ref{fig:card_sirSSPL45} we did not perform measurements with distances greater than $15$ cm, we cannot conclude if the SIR curves would exhibit the similar behavior as the SIR curves from Figures~\ref{fig:omni_sirSSPL} and~\ref{fig:card_sirSSPL}, i.e. a deep at certain distance followed by a small increase towards a high limit of the SIR. 

The SIR values obtained with the placement of the cardioid microphones with an angle are almost three times the values of the SIR that were obtained with the other two cases. This clearly indicates that placing a cardioid microphone with an angle against the central axis of the noise results in better performance of the close miking technique. These values of SIR in the corresponding peaks are almost three times the peak SIR values from the rest two cases of microphone types and angles of placement. This clearly indicates the outperformance of the cardioid microphones placed with an angle versus the cardioid microphone placed without an angle and the omni-directional microphone cases. Finally, between the two different angular placements of the cardioid microphones, there is not any notable difference with the current experimental setup. 

\section{Conclusions}\label{sec:conclusions}
\vspace{-6pt}
The work at hand performed a quantitative analysis of the source separation capabilities of the close miking technique. Since this technique is a mechanical source separation method, the present work applies a quantitative analysis of the actual close miking technique. This analysis is performed with two different microphone types, three different angular placements of the microphones, $12$ different distances between the microphone and the source, three different source SPL, and, finally, under five different noise SPL values. 

The results obtained clearly indicate that the best performance of  close miking is achieved when the microphone has a cardioid lobe, placed with an angle of $30$ or $45$ degrees with respect to the central axis of the source and in distance of around $12$ cm. 

Future measurements and studies could, potentially, show the effect of the height of the microphone in the close miking technique. Finally, there would increased interest in a subjective evaluation of the quality of the source separation with close miking with different types of microphones and different angular placements of the microphones with respect to the source. 
\section{Acknowledgements}
The authors would like to thank the Department of Technology of Sound and Musical Instruments, Technological Educational Institute of Ionian Islands, for providing the equipment for the measurements. Part of the research leading to these results has received funding from: i) the European Research Council under the European Union's H2020 Framework Programme through ERC Grant Agreement 637422 EVERYSOUND, and ii) the European Union's H2020 Framework Programme (H2020-MSCA-ITN-2014) under grant agreement no 642685 MacSeNet.


\end{document}